\documentclass[%
superscriptaddress,
 amsmath,amssymb,
 aps,prl,
 twocolumn,
]{revtex4-2}
\usepackage{color}
\usepackage{graphicx}% Include figure files
\usepackage{dcolumn}% Align table columns on decimal point
\usepackage{bm}% bold math

\begin{document}

%\preprint{APS/123-QED}

\title{Double Asymptotic Structures of Topologically Interlocked Molecules}%

\author{Jiang-Tao Li}
\thanks{These authors contributed equally to this work.}
\author{Fang Gu}
\thanks{These authors contributed equally to this work.}
\author{Ning Yao}
\affiliation{College of Chemistry and Environmental Science, Hebei University, Baoding, Hebei (071002), China.}
\author{Hai-Jun Wang}%
\email{whj@hbu.edu.cn}
\affiliation{College of Chemistry and Environmental Science, Hebei University, Baoding, Hebei (071002), China.}
\author{Qi Liao}%
\email{qiliao@iccas.ac.cn}
\affiliation{Institute of Chemistry, Chinese Academy of Sciences, Beijing (100190), China.}

\date{\today}% It is always \today, today,

\begin{abstract}
The mean square size of topologically interlocked molecules (TIMs) is presented as a linear combination of contributions from the backbone and subcomponents. Using scaling analyses and extensive molecular dynamics simulations of polycatenanes, as a typical example of TIMs, we show that the effective exponent $\nu(m)$ for the size dependence of the backbone on the monomer number of subcomponent $m$ is asymptotic to a value $\nu$ (approximately 0.588 in good solvents) with a correction of $m^{-0.47}$, which is the same as for the covalently linked polymer. However, the effective exponent for the size dependence of subcomponents on $m$ is asymptotic to the same value $\nu$ but with a new correction of $m^{-1.0}$.
The different corrections to the scaling on the backbone and subcomponent structure induce a surprising double asymptotic behavior for the architecture of the TIMs.
The scaling model that takes into account the double asymptotic behavior is in good quantitative agreement with the simulation result that the effective exponent for the size dependence of TIMs on $m$ increases with the subcomponent number $n$.
The full scaling functional form of the size dependence on $m$ and $n$ for polycatenanes in a good solvent is well described by a simple sum of two limiting behaviors with different corrections.
\end{abstract}

\maketitle

%\tableofcontents

%\newpage
Mechanically interlocked molecules consist of subcomponents connected together without
being covalently linked\cite{Hart2021,Wasserman1960,Sauvage2017,Feringa2017,Stoddart2017,
Sluysmans2019}.
These molecules can be divided into two classes based on the different lock architectures. One class of mechanically interlocked molecules, rotaxanes for example, is interlocked by the excluded-volume interaction and detachable from the topological point of view. The second class is topologically entangled by the unbroken covalent bonds between subcomponents. In this letter, we define the second class of mechanically interlocked molecules containing subcomponents as topologically interlocked molecules (TIMs), as shown in Figure 1. An isolated molecular knot would not be regarded as TIMs under this definition, because there are no subcomponents in the knot.

The linkage of two ring chains or catenanes has been investigated by experiments\cite{Hudson1967,Clayton1967,Krajina2018}, theoretical analysis, and computer simulations\cite{Xiong2012}. The square radius of gyration of the TIMs with two identical subcomponents $R_\text G^{2}$ is the sum of the mean square radius of gyration of the subcomponent, $R_\text g^{2}$, and a quarter of the mean square distance between the centers of the subcomponent mass, $d^{2}$. More generally, the size dependence of TIMs containing $n$ subcomponents can be calculated exactly by the following expression:
\begin{equation}
R_\text G^{2}(n)=\sum_{i=1}^{n}f_{i} R_{i}^2+
\frac{1}{2}\sum_{i=1}^{n}\sum_{j=1}^{n}f_{i}f_{j}d_{ij}^2,
\end{equation}
where $f_{i}$ and $R_\text g^{2}(i)$ are the mass fraction and the square radius of gyration of the $i$-th subcomponent, respectively, and $d_{ij}$ is the distance between the mass centers of the $i$-th and $j$-th subcomponents. On the right-hand side of Eq. (1), the first term represents the contribution of the subcomponents, and the second term represents the contribution of the TIMs backbone characterized by the centers of mass of the subcomponents.

\begin{figure}[th]
\includegraphics[width=8.8cm]{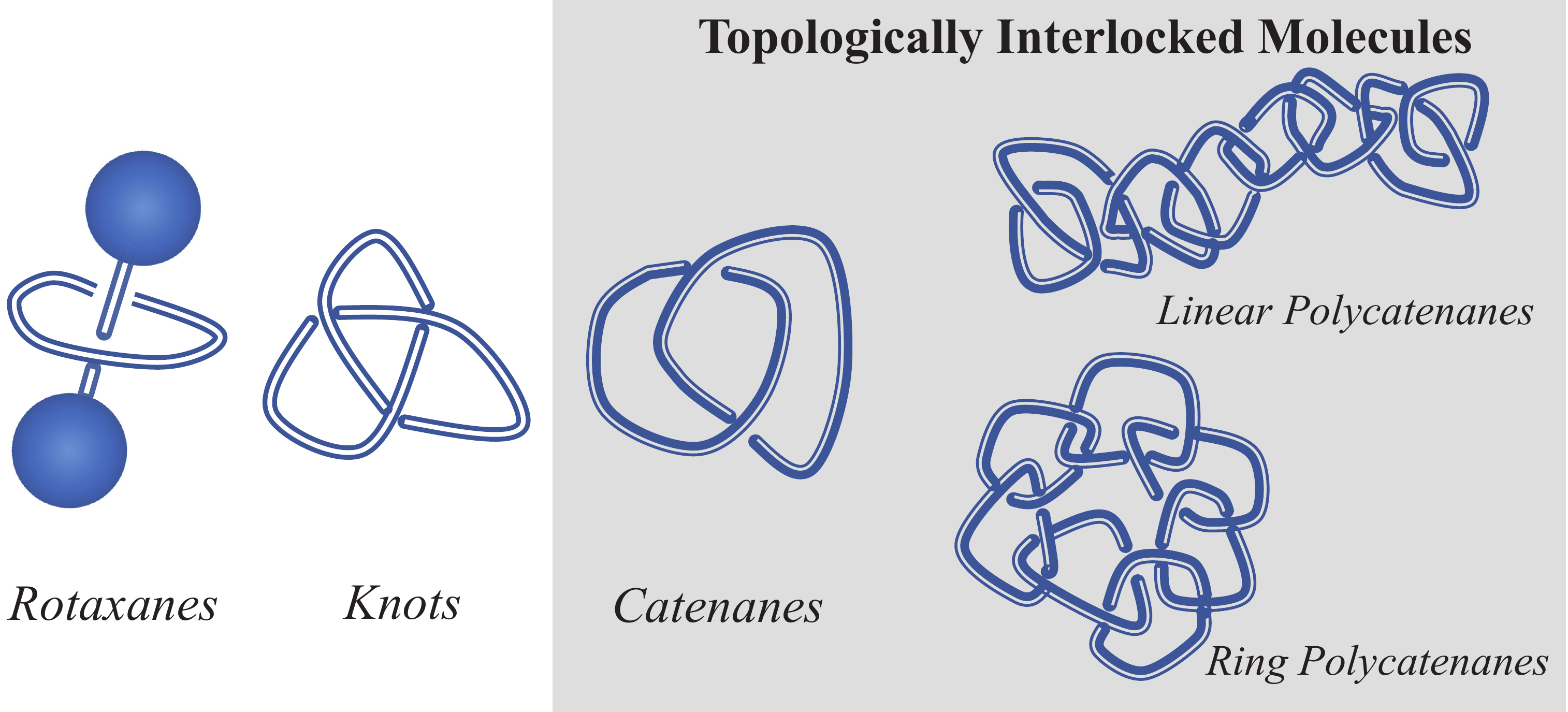}
\caption{Schematic illustration of TIMs under study, which are defined as the molecules topologically entangled between subcomponents.}
\label{fig:cartoon}
\end{figure}

For the case of identical $n$ subcomponents with $m$ monomers each, such as polycatenanes\cite{Weidmann1999,Pakula1999,Wu2017,Rauscher2018}, Eq. (1) becomes
\begin{equation}
{R_{\rm{G}}^2(n,m)}=\langle{R_{\rm{g}}^2(m)}\rangle +F(n)\langle d^2(m)\rangle,
\end{equation}
where $\langle d^2(m)\rangle$ is the mean square distance between the centers of mass of neighboring subcomponents, and $F(n)$ is a function of the subcomponent number $n$. Eq. (2) states that the mean square size of TIMs is a linear combination of the mean square size of the subcomponents and the backbone of TIMs. Eq. (2) also provides a universal strategy to predict the size dependence of TIMs by separately analyzing the dimension of the subcomponents, backbone, and distance between neighboring subcomponents. Note that $F(n)$ depends on the interlocked architecture of TIMs, as shown in Fig. 1. Clearly, as the subcomponent number increases, the contribution of the backbone dominates the size of the TIMs. For the cases of linear and ring polycatenanes, the scaling model of thermal blobs can be applied and leads to $F(n)\sim {n^{2\nu}}$ when the subcomponent size is larger than the size of the thermal blob\cite{Gennes1979,Rubinstain2003}.

To check the scaling prediction of the size dependence on $n$, we perform molecular dynamics simulations of polycatenanes with $n$ varying from 64 to 256 and $m$ varying from 16 to 256 under good solvent conditions. The simulation results of the end-to-end distance $R_\text{end}$ and radius of gyration $R_\text{g}$ of polycatenanes with m=16 are shown in Fig. 2. The simulation model and details have been described in Ref. [10]. Figure 2a shows the internal distance of the sub-ring center of mass $R_\text{end,cm}(|i-j|)$ and radius of gyration $R_\text{G}(|i-j|)$ for polycatenanes of $m=16$ with a different sub-ring number obtained from our simulation, where $i$ and $j$ are the indices of the sub-ring.

\begin{figure}[th]
\includegraphics[width=8.8cm]{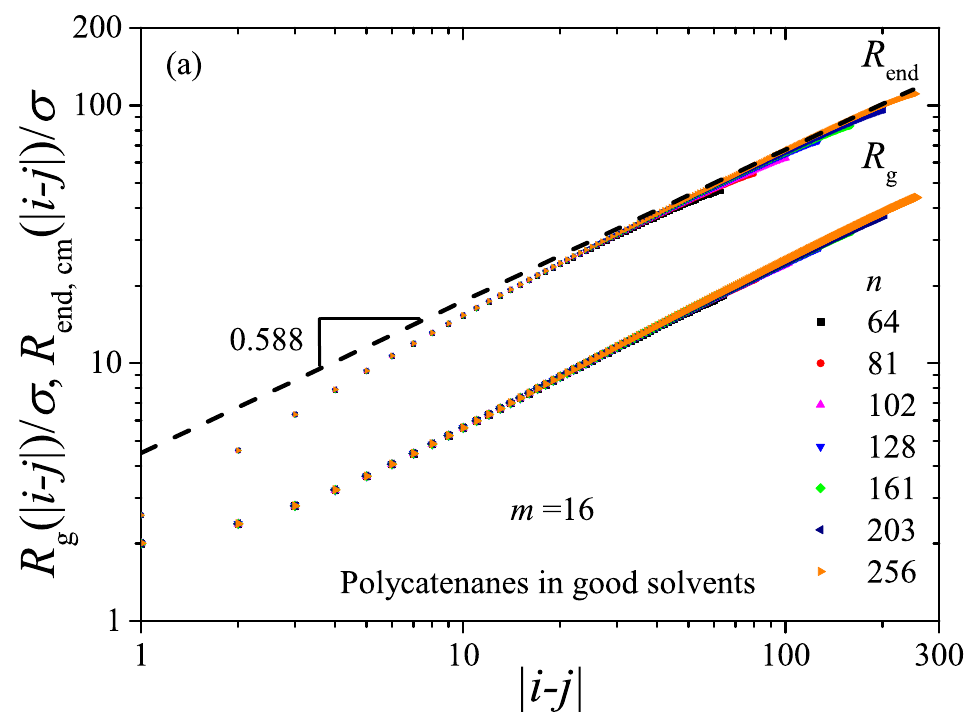}
\includegraphics[width=8.8cm]{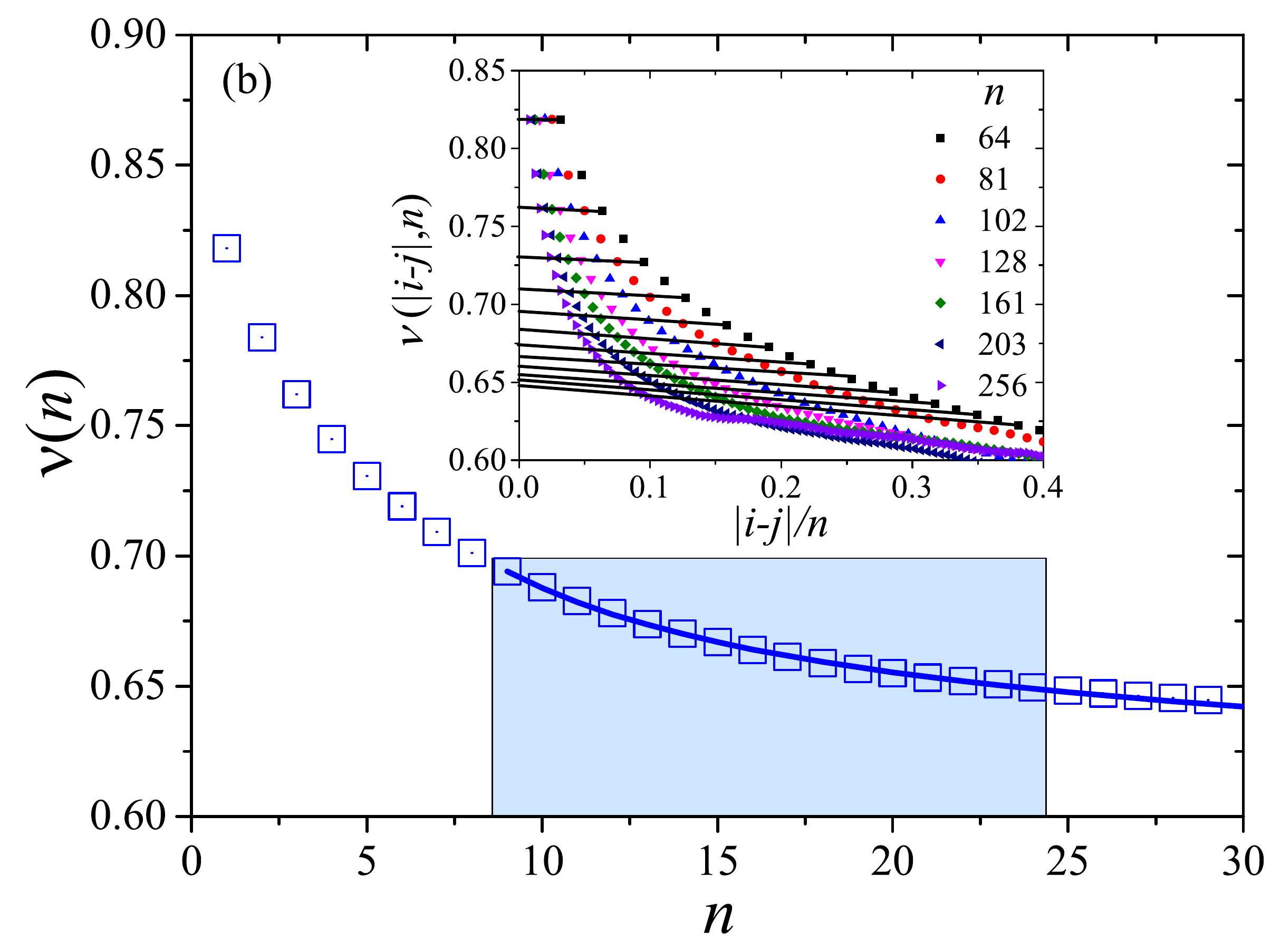}
\caption{(a) Size dependence of polycatenanes on the inter-molecule sub-ring index $|i-j|$ and $n$, where $\sigma$ is the distance unit of the Lennard-Jones interaction. (b) The exponent and correction are fitted by Eq. (3). The effective exponents $\nu(|i-j|,\infty)$ for the size on $|i-j|$ are extrapolated by the internal distance of polycatenanes $|i-j|$ to infinite $n$ (shown in inset). Note that $\nu(|i-j|,\infty)$ has been rewritten as $\nu(n)$ by the change of variable $|i-j|$ to $n$ for consistency with Eqs. (3) and (4). The blue area shows the fitting range of $n$.}
\label{fig:cartoon}
\end{figure}

The scaling behavior with the correction term of the size dependence of polycatenanes on $n$, which involves a coefficient $B$, can be given by\cite{Cloizeaus1990,Guillou1980}
\begin{equation}
F(n)\sim {n^{2\nu}}(1+Bn^{-\Delta}),
\end{equation}
where $\Delta$ is the correction exponent. By defining the effective exponent $\nu (n)=\frac{1}{2}\frac{\partial \ln F(n)}{\partial \ln n}$, we have
\begin{equation}
\nu (n)=\nu-\frac{B\Delta}{B+n^{\Delta}}.
\end{equation}
Here, we develop the extrapolation technique suggested by Havlin et al. \cite{Havlin1983} to extrapolate the sub-ring number to infinity by analysis of the internal distance, as shown in the inset of Fig. 2b. The effective exponent for the inter-ring distance is defined as $\nu (|i-j|,n)=\frac{1}{2}\frac{\partial \ln \langle R_\text{end,cm}^2(|i-j|)\rangle}{\partial \ln |i-j|}$. By the extrapolation and change of variable, the simulation results of $\nu (n)$ shown in Fig. 2b are fitted by Eq. (4). The best-fitting parameters are given as $\nu=0.589\pm0.006$, $B=-0.513\pm0.010$, and $\Delta=0.514\pm0.053$. The results are in good agreement with the field theory calculation of $\nu=0.588$ and $\Delta=0.47$ for the covalently linked polymer chain in good solvents\cite{Guillou1980}.

\begin{figure}
\includegraphics[width=8.8cm]{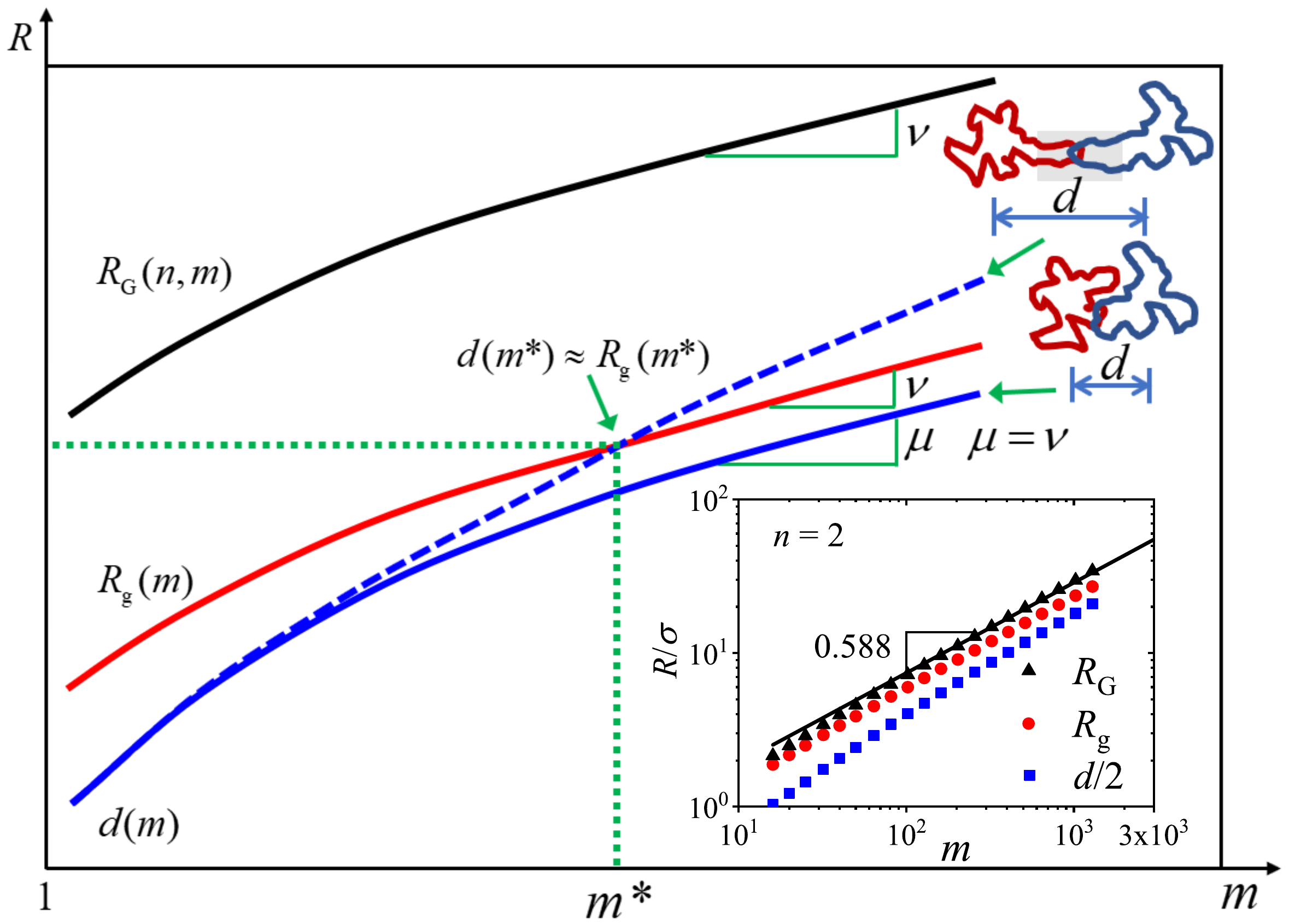}
\caption{Scaling of size of TIMs and corrections in log-log scales. The inset shows the dependences of $ R_{\rm{G}}$, $R_{\rm{g}}$, and $\frac{d}{2}$ on $m$ of the two linked rings by using molecular dynamics simulations. The solid line with a slope of 0.588 is a guide for the eye.}
\label{fig 3}
\end{figure}

After the estimation of the effective exponent and correction of the subcomponent number, we may further determine the effective exponent of the monomer number of subcomponent $m$ in Eq. (2). For linear and ring polycatenanes, the scaling model of the thermal blob leads to $R_\text{g}(m)\sim m^{\nu}b$, where $b$ is the average bond length. Furthermore, one can assume that $d(m) \sim m^{\mu}b$. Note that all sub-rings are finitely stretched, and therefore we must have $\nu\geq \mu$, otherwise these sub-rings would be infinitely stretched when $m$ is larger than $m^{*}$ which is defined as $d(m^*) \approx R_{\rm{g}}(m^*)$, as illustrated by the blue dashed line in Fig. 3. This means that the scaling of $d(m)$  will be asymptotic to the same scaling of $ R_{\rm{g}}(m)$
due to the finite extension of the sub-rings, and hence the scaling of the total size of polycatenanes $ R_{\rm{G}}(m)$ also will be asymptotic to the same exponent of $m$
as
$R_{\rm{G}}(m) \sim R_{\rm{g}}(m) \sim d(m) \sim m^\nu$.

For the quantities $R_{\rm{g}}(m)$ and $d(m)$ under study, the expressions of the scaling law with corrections are
\begin{eqnarray}
\left\{
\begin{array}{l}
R_{\rm{g}}(m) \sim {m^{\nu}}(1+B_{1}m^{-\Delta_{1}}), \vspace{1ex}\\
d(m)\sim  {m^{\nu}}(1+B_{2}m^{-\Delta_{2}}).
\end{array}
\right.
\end{eqnarray}

The scaling predictions and corrections are checked by our molecular dynamics simulations, and the results for catenanes are shown in the inset of Fig. 3. By defining the effective exponents as
$\nu_{\text g}(m)=\frac{1}{2}\frac{\partial \ln \langle R_{\rm{g}}^2(m)\rangle}{\partial \ln m}$
and  $\nu_{\text d}(m)=\frac{1}{2}\frac{\partial \ln \langle d^2(m)\rangle}{\partial \ln m}$, they can be written in the form
\begin{eqnarray}
\left\{
\begin{array}{l}
\nu_{\text g}(m)=\nu-\frac{B_{1}\Delta_{1}}{B_{1}+m^{\Delta_{1}}},\vspace{1ex}\\
\nu_{\text d}(m)=\nu-\frac{B_{2}\Delta_{2}}{B_{2}+m^{\Delta_{2}}}.
\end{array}
\right.
\end{eqnarray}
The effective exponents for  the two linked rings and three interlocked rings obtained by using molecular dynamics simulations are given in Fig. 4. The results for $\nu_{\text g}(m)$ and $\nu_{\text d}(m)$ shown in Fig. 4 are fitted by Eq. (6), and the parameters are given in Table 1.

\begin{figure}[th]
\includegraphics[width=8.8cm]{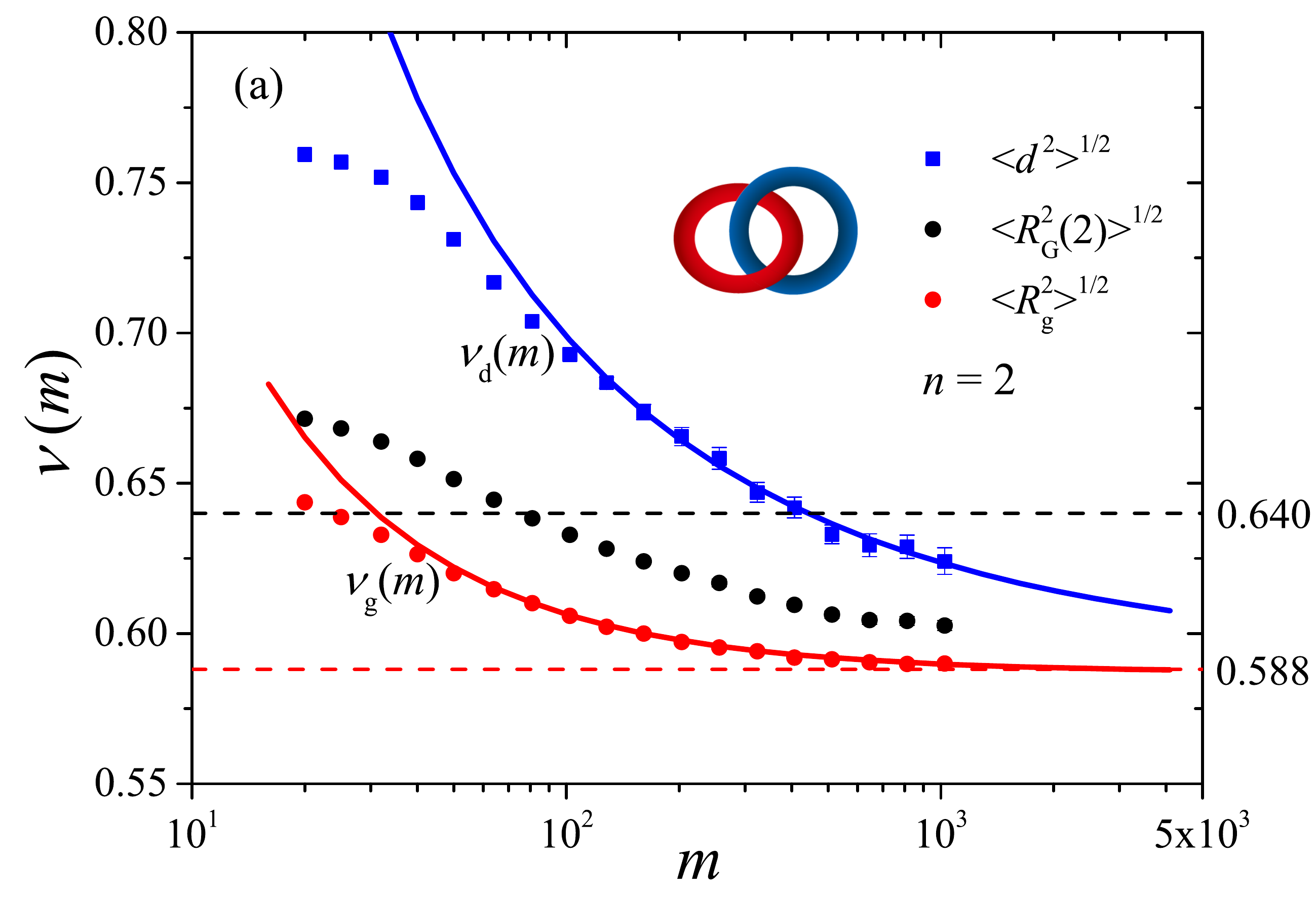}
\includegraphics[width=8.8cm]{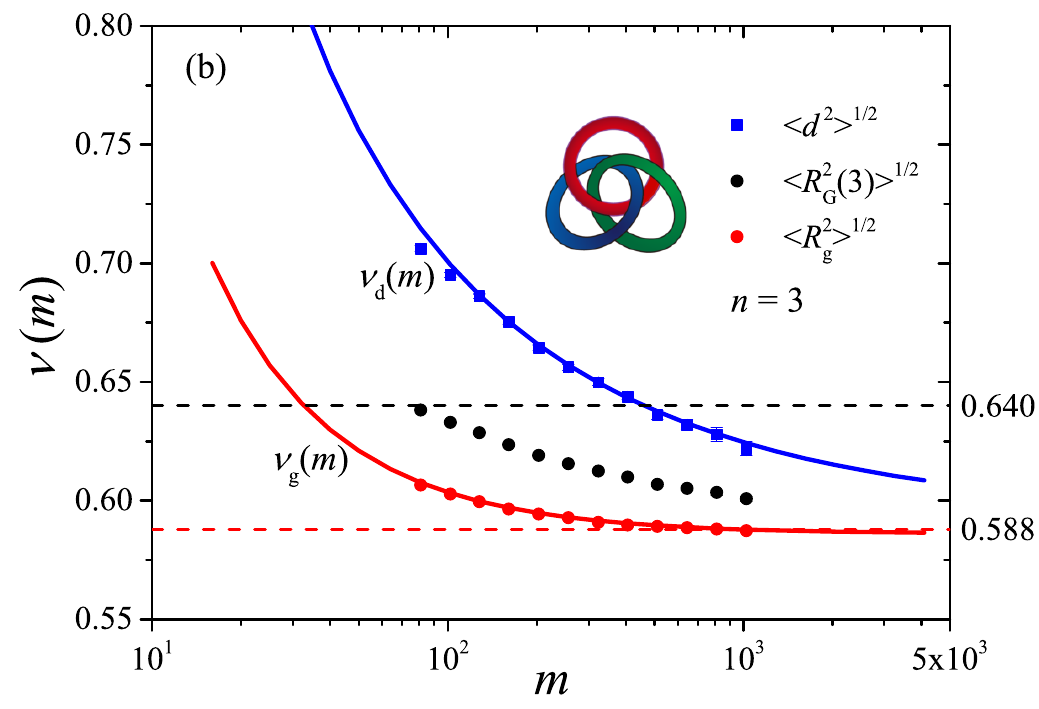}
\caption{Effective scaling exponent as a function of $m$ for (a) two linked rings, and (b) three interlocked rings. The dashed lines at positions $\nu(m)=$ 0.588 and 0.640 are guides for the eye. The solid lines are the best-fitting curves of Eq. (6). See more details in the text.}
\label{fig 4}
\end{figure}

\begin{table*}
\caption{\label{tab:table1}Fitting results of molecular dynamics simulation data for the two linked rings and three interlocked rings.}
\begin{ruledtabular}
 \begin{tabular}{ccccccc}
 &\multicolumn{3}{c}{$\langle R_{\rm{g}}^2(m)\rangle^{\frac{1}{2}}$}&\multicolumn{3}{c}{$\langle d^2(m)\rangle^{\frac{1}{2}}$}\\  $n$&$\nu$ &$\Delta_{1}$&$B_{1}$ &$\nu$ &$\Delta_{2}$ &$B_{2}$
 \\ \hline
  2&0.587$\pm$0.001 &0.93$\pm$0.11&-1.44$\pm$0.57 & 0.591$\pm$0.003 & 0.46$\pm$0.10 &-1.57$\pm$0.54\\  3&0.586$\pm$0.001 &0.97$\pm$0.12&-1.55$\pm$0.64 & 0.592$\pm$0.007 & 0.46$\pm$0.07 &-1.59$\pm$0.24\\
 \end{tabular}
\end{ruledtabular}
\end{table*}

Here, we surprisingly find that the sub-ring and backbone exhibit different asymptotic behaviors as the monomer number $m$ varies, as shown in Fig. 4 and Table 1. The effective exponent for the dependence of $\langle d^2(m)\rangle^{\frac{1}{2}}$ on $m$ is asymptotic to the value $\nu=0.588$ by a correction of $m^{-0.46\pm0.10}$, which is same as for the covalently linked polymer in good solvents[17-19]. However, the effective exponent for the size dependence of $\langle R_{\rm{g}}^2(m)\rangle^{\frac{1}{2}}$ on $m$ is asymptotic to 0.588 but with a new correction of $m^{-1.0}$, which has not been reported up to now to the best of our knowledge. Such double asymptotic structures can be characterized more clearly by the characteristic ratio defined by $C_{m}\equiv\frac{\langle \chi^2(m)\rangle}{m^{2\nu}b^2}$ as
\begin{equation}
{C_\infty}-{C_m} \sim {m^{-\Delta_i}},
\end{equation}
where $\chi^2(m)$ is $R_{\text{g}}^2(m)$ for $i=1$ and $d^2(m)$ for $i=2$. The simulation results of the dependence of ${C_\infty } - {C_m}$ on the monomer number of ring $m$ for two linked rings and three interlocked rings are shown in Fig. 5. The results clearly confirm the double asymptotic structures on the monomer number $m$.

Substituting Eqs. (3) and (5) into Eq. (2), the full scaling functional form of the size dependence on $m$ and $n$ for linear and ring polycatenanes in good solvents can be well described by
\begin{eqnarray}
\frac{{R_{\rm{G}}^2(n,m)}}{{m^{2\nu}}b^2} &\cong& A_{1}(1+2B_{1}m^{-\Delta_{1}})+A_{2}{(n)^{2\nu}}\nonumber\\
&&\times(1+2Bn^{-\Delta})(1+2B_{2}m^{-\Delta_{2}}),
\end{eqnarray}
where $\Delta_{1}\cong1.0$,
$\Delta\cong\Delta_{2}\cong 0.47$, and $A_{1}$, $A_{2}$, $B$, $B_{1}$, and $B_{2}$ are constants depending on the chemical structure of polycatenanes.
According to the function form, the effective exponent of TIMs is predicted to shift from the lower limit boundary given by the subcomponents to the upper limit boundary given by the backbone as the number of subcomponents increases, as shown in Fig. 4 by the black points.

\begin{figure}[th]
\includegraphics[width=8.8cm]{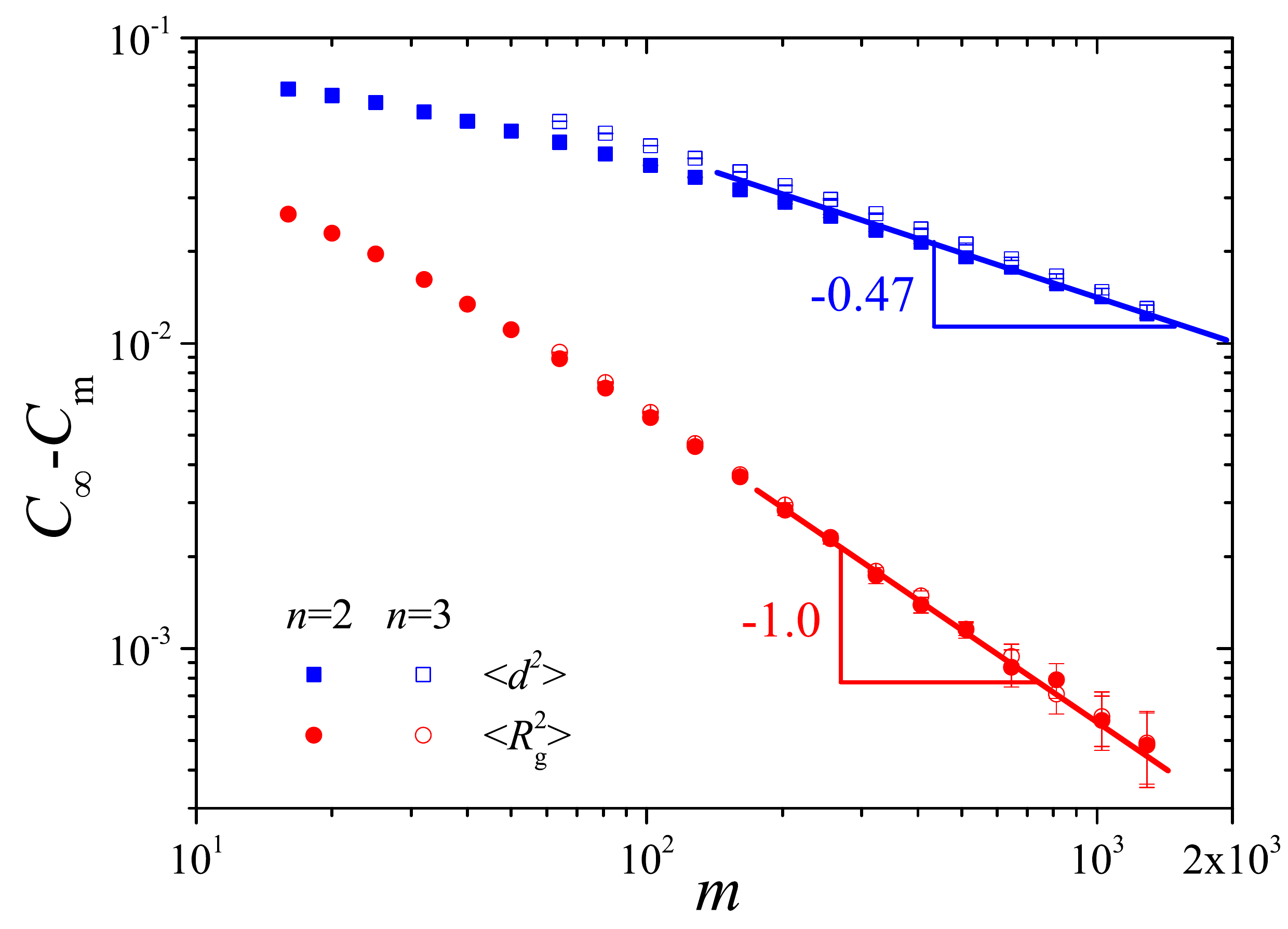}
\caption{Dependence of $C_{\infty}-{C_m}$ on $m$ for two linked rings and three interlocked rings. The solid lines with a slope of $-0.47$ and $-1.0$ are the guide to the eye.}
\label{fig 5}
\end{figure}

Recently, Dehaghani et al\cite{Dehaghani2020} observed that the exponent for the size dependence of polycatenanes on $m$ is approximately $0.64$ in good solvents, which is larger than 0.588. Note that in their work the largest monomer number of sub-ring $m$ is approximately 256, and our present simulation results, in which the largest monomer number extends to 1024, confirm that the effective exponent is asymptotic to 0.588 in good solvents.
The effective exponent of approximately 0.64 reported in Ref. [20] is due to the double asymptotic structure of polycatenanes, which shifts to a larger value as the subcomponent number $n$ increases at finite $m$ around 256.

In summary, we propose a universal way to predict the size dependence of TIMs on the architectural parameters by applying the scaling law with corrections and simulations. By using extrapolation techniques to extrapolate the subcomponent number $n$ and the monomer number $m$ to infinity, we find double asymptotic structures that depend on the monomer number of subcomponent in the TIMs. As a typical example of TIMs, the full scaling functional form of the size dependence on $m$ and $n$ for linear polycatenanes in a good solvent has been presented. Many other TIMs can be studied using a similar type of analysis. The scaling functional form of the size dependence can also be constructed by a similar protocol as shown in this letter.
~\\
\begin{acknowledgments}
This work is supported by the National Natural Science Foundation of China under Grant 21574139, 21973103. We are also grateful to the High-Performance Computing Center of Hebei University.
\end{acknowledgments}

%\newpage

\nocite{*}

\end{document}